\documentclass[a4paper]{amsart}

\RequirePackage{amsmath}
\RequirePackage{bm}
\RequirePackage{amssymb}
\RequirePackage{upref}
\RequirePackage{amsthm}
\RequirePackage{enumerate}
\RequirePackage{pb-diagram}
\RequirePackage{amsfonts}
\RequirePackage[mathscr]{eucal}
\RequirePackage{verbatim}
\RequirePackage{xr}
\RequirePackage{graphicx}
\usepackage{calc}
\usepackage{xspace}
\RequirePackage{color}
\RequirePackage{ifthen}

\newcommand{\cf}{cf.\@\xspace}
\newcommand{\resp}{resp.\@\xspace}

\newcommand{\al}{\alpha}
\newcommand{\bet}{\beta}
\newcommand{\ga}{\gamma}

\newcommand{\e}{\epsilon}

\newcommand{\f}{\varphi}
\newcommand{\h}{\eta}

\newcommand{\lam}{\lambda}
\newcommand{\m}{\mu}

\newcommand{\om}{\omega}

\newcommand{\C}{\varGamma}

\newcommand{\F}{\varPhi}
\newcommand{\Lam}{\varLambda}
\newcommand{\Om}{\varOmega}

\newcommand{\fv}[2]{#1\hspace{0pt}_{|_{#2}}}

\newcommand{\so}{{\mc S_0}}

\newcommand{\const}{\tup{const}}

\newcommand{\msp[1]}[1]{\mspace{#1mu}}

\newcommand{\R}[1][n+1]{{\protect\mathbb R}^{#1}}

\newcommand{\Cc}{{\protect\mathbb C}}
\newcommand{\K}{{\protect\mathbb K}}
\newcommand{\N}{{\protect\mathbb N}}

\newcommand{\eR}{\stackrel{\lower1ex \hbox{\rule{6.5pt}{0.5pt}}}{\msp[3]\R[]}}
\newcommand{\eN}{\stackrel{\lower1ex \hbox{\rule{6.5pt}{0.5pt}}}{\msp[1]\N}}
\newcommand{\eO}{\stackrel{\lower1ex \hbox{\rule{6pt}{0.5pt}}}{\msc O}}
\newcommand{\mf}[1]{\mathfrak {#1}}

\DeclareMathOperator{\supp}{supp}
\DeclareMathOperator{\id}{id}

\DeclareMathOperator{\Ad}{Ad}

\DeclareMathOperator{\tr}{tr}

\DeclareMathOperator{\diag}{diag}

\newcommand\ra{\rightarrow}

\newcommand\pde[2]{\frac {\partial#1}{\partial#2}}
 
\newcommand\df[2]{\frac {d#1}{d#2}}

\newcommand{\un}{\infty}
\newcommand{\A}{\forall}

\newcommand{\uu}{\cup}
\newcommand{\ii}{\cap}
\newcommand{\uuu}{\bigcup}

\newcommand{\uud}{ \stackrel{\lower 1ex \hbox {.}}{\uu}}
\newcommand{\uuud}[1]{ \stackrel{\lower 1ex \hbox {.}}{\uuu_{#1}}}
\newcommand\su{\subset}
\newcommand\Su{\Subset}

\newcommand{\sminus}[1][28]{\raise 0.#1ex\hbox{$\scriptstyle\setminus$}}

\newcommand{\wed}{\wedge}

\newcommand{\abs}[1]{\lvert#1\rvert}

\newcommand{\spd}[2]{\protect\langle #1,#2\protect\rangle}

\newcommand{\tit}{\textit}

\newcommand{\tup}{\textup}

\newcommand{\mc}{\protect\mathcal}
\newcommand{\msc}{\protect\mathscr}

\providecommand{\bysame}{\makebox[3em]{\hrulefill}\thinspace}

\newcommand{\bt}{\begin{thm}}
\newcommand{\bl}{\begin{lem}}
\newcommand{\bc}{\begin{cor}}
\newcommand{\bd}{\begin{definition}}
\newcommand{\bpp}{\begin{prop}}
\newcommand{\br}{\begin{rem}}
\newcommand{\bn}{\begin{note}}
\newcommand{\be}{\begin{ex}}
\newcommand{\bes}{\begin{exs}}
\newcommand{\bb}{\begin{example}}
\newcommand{\bbs}{\begin{examples}}
\newcommand{\ba}{\begin{axiom}}
\newcommand{\bas}{\begin{assumption}}

\newcommand{\et}{\end{thm}}
\newcommand{\el}{\end{lem}}
\newcommand{\ec}{\end{cor}}
\newcommand{\ed}{\end{definition}}
\newcommand{\epp}{\end{prop}}
\newcommand{\er}{\end{rem}}
\newcommand{\en}{\end{note}}
\newcommand{\ee}{\end{ex}}
\newcommand{\ees}{\end{exs}}
\newcommand{\eb}{\end{example}}
\newcommand{\ebs}{\end{examples}}
\newcommand{\ea}{\end{axiom}}
\newcommand{\eas}{\end{assumption}}

\newcommand{\bp}{\begin{proof}}
\newcommand{\ep}{\end{proof}}
\newcommand{\eps}{\renewcommand{\qed}{}\end{proof}}

\newcommand{\bal}{\begin{align}}

\newcommand{\bi}[1][1.]{\begin{enumerate}[\upshape #1]}
\newcommand{\bia}[1][(1)]{\begin{enumerate}[\upshape #1]}
\newcommand{\bin}[1][1]{\begin{enumerate}[\upshape\bfseries #1]}
\newcommand{\bir}[1][(i)]{\begin{enumerate}[\upshape #1]}
\newcommand{\bic}[1][(i)]{\begin{enumerate}[\upshape\hspace{2\cma}#1]}
\newcommand{\bis}[2][1.]{\begin{enumerate}[\upshape\hspace{#2\parindent}#1]}
\newcommand{\ei}{\end{enumerate}}

\newcommand\ndots{\raise 0.47ex \hbox {,}\hskip0.06em\cdots %
     \raise 0.47ex \hbox {,}\hskip0.06em} 

\newcommand{\q}{\quad}
\newcommand{\qq}{\qquad}

\newcommand{\hp}{\hphantom}

\newcommand\nd{\noindent}

\newskip\Csmallskipamount                                                
\Csmallskipamount=\smallskipamount
\newskip\Cmedskipamount
\Cmedskipamount=\medskipamount
\newskip\Cbigskipamount
\Cbigskipamount=\bigskipamount

\newcommand\cvs{\vspace\Csmallskipamount}   
\newcommand\cvm{\vspace\Cmedskipamount}

\newskip\csa
\csa=\smallskipamount

\newskip\cma
\cma=\medskipamount

\newskip\cba
\cba=\bigskipamount

\newdimen\spt
\newcommand\citem{\cvs\advance\itemno by
1{(\romannumeral\the\itemno})\hskip3pt}
\newcommand{\bitem}{\cvm\nd\advance\itemno by
1{\bf\the\itemno}\hspace{\cma}}

\newcount\itemno
\newcommand{\lae}[1]{\label{E:#1}}
\newcommand{\lat}[1]{\label{T:#1}}
\newcommand{\lal}[1]{\label{L:#1}}
\newcommand{\lad}[1]{\label{D:#1}}

\newcommand{\rl}[1]{Lemma~\ref{L:#1}}

\newcommand{\re}[1]{\eqref{E:#1}}
\newcommand{\frt}[1]{Theorem~\ref{T:#1} on page~\tup{\pageref{T:#1}}}

\newcommand{\frd}[1]{Definition~\ref{D:#1} on page~\tup{\pageref{D:#1}}}
\newcommand{\fre}[1]{\eqref{E:#1} on page~\tup{\pageref{E:#1}}}

\newskip\thmskip
\thmskip=\parindent

\newskip\hsk
\newenvironment{hinw}{\labelsep=0pt\begin{list}{}{\labelsep=0pt\itemindent=0pt\labelwidth=0pt\leftmargin=\parindent\rightmargin=0pt\partopsep=\cba}%
\item\it\nopagebreak\nopagebreak}%
{\end{list}}

\newcommand\bh{\begin{hinw}}
\newcommand{\eh}{\end{hinw}}

\newtheoremstyle{normal}
  {\cba}
  {\cba}
  {}
  {\thmskip}
  {\bfseries}
  {.}
  {\hsk}
  {}

\newtheoremstyle{abschnitt}
  {\cba}
  {\cba}
  {}
  {\thmskip}
  {\bfseries}
  {.}
  {\hsk}
  {}

\newtheoremstyle{italic}
  {\cba}
  {\cba}
  {\itshape}
  {\thmskip}
  {\bfseries}
  {.}
  {\hsk}
  {}

\newtheoremstyle{aufgaben}
  {\cba}
  {\cba}
  {}
  {}
  {\normalsize\bfseries}
  {.}
  {\hsk}
  {}

\newtheoremstyle{break}
  {\cba}
  {\cba}
  {\itshape}
  {}
  {\bfseries}
  {.}
  {\newline}
  {}

\swapnumbers
\theoremstyle{italic}
\newtheorem{thm}[subsection]{Theorem}
\newtheorem{lem}[subsection]{Lemma}
\newtheorem{prop}[subsection]{Proposition}
\newtheorem{cor}[subsection]{Corollary}

\theoremstyle{normal}
\newtheorem{rem}[subsection]{Remark}
\newtheorem{definition}[subsection]{Definition}
\newtheorem{example}[subsection]{Example}
\newtheorem{examples}[subsection]{Examples}
\newtheorem{ex}[subsection]{Exercise}
\newtheorem{note}[subsection]{}
\newtheorem{axiom}[subsection]{Axiom}
\newtheorem{assumption}[subsection]{Assumption}

\theoremstyle{aufgaben}
\newtheorem{exs}[subsection]{Exercises}

\swapnumbers

\numberwithin{equation}{section}
\numberwithin{figure}{section}

\newenvironment{textequation}[1][0.8]
{\begin{equation}
\begin{aligned}
\begin{minipage}{#1\linewidth}}
{\end{minipage}
\end{aligned}
\end{equation}
\ignorespacesafterend}

\newcommand{\btext}{\begin{textequation}}
\newcommand{\etext}{\end{textequation}}

\def\hinweis{\@startsection{subsection}{2}%
 \z@{0.7\linespacing\@plus 0.5\linespacing}{0.7\linespacing}%
{\normalfont\itshape\indent}}

\newcounter{hours}\newcounter{minutes}
\newcommand{\printtime}{%
\setcounter{hours}{\time/60}%
\setcounter{minutes}{\time-\value{hours}*60}%
\ifthenelse{\value{minutes}<10}{\thehours :0\theminutes}{\thehours:\theminutes}}

\usepackage[german,english]{babel}
\usepackage{graphicx}
\RequirePackage{amsmath}
\RequirePackage{bm}
\RequirePackage{amssymb}
\RequirePackage{upref}
\RequirePackage{amsthm}
\RequirePackage{enumerate}
\RequirePackage{pb-diagram}
\RequirePackage{amsfonts}
\RequirePackage[mathscr]{eucal}
\RequirePackage{verbatim}
\RequirePackage{xr}
\RequirePackage{graphicx}
\usepackage{calc}
\usepackage{xspace}

\makeatletter
\RequirePackage{color}
\newcommand{\ann}[1]{\renewcommand{\@makefnmark}{\mbox{$^{\color{red}{\@thefnmark}}$}}%
\footnote {#1}}
\makeatother

\RequirePackage{upref}
\RequirePackage{amsthm}
\RequirePackage{enumerate}
\usepackage[mathscr]{eucal}

\usepackage{xr-hyper}

\usepackage{calc}

\newlength{\oddsidemarginlength}
\newlength{\topmarginlength}

\newcounter{numberoflines}
\newcounter{tempcc}
\oddsidemargin=\oddsidemarginlength
\evensidemargin=\oddsidemargin
\topmargin=\topmarginlength
\usepackage[colorlinks=true,linkcolor=blue,citecolor=blue,urlcolor=blue]{hyperref}  

\begin{document}

\flushbottom


\title[A unified quantum theory I]{A unified quantum theory I: gravity interacting with a Yang-Mills field}

\author{Claus Gerhardt}
\address{Ruprecht-Karls-Universit\"at, Institut f\"ur Angewandte Mathematik,
Im Neuenheimer Feld 294, 69120 Heidelberg, Germany}
\email{\href{mailto:gerhardt@math.uni-heidelberg.de}{gerhardt@math.uni-heidelberg.de}}
\urladdr{\href{http://www.math.uni-heidelberg.de/studinfo/gerhardt/}{http://www.math.uni-heidelberg.de/studinfo/gerhardt/}}

%
\subjclass[2000]{83,83C,83C45}
\keywords{globally hyperbolic Lorentzian manifold, quantum gravity, Yang-Mills field, strong interaction, unification, unified quantum theory}

\date{\today}
%


\begin{abstract} 
Using the results and techniques of a previous paper where we proved the quantization of gravity we extend the former result by adding a Yang-Mills functional and a Higgs term to the Einstein-Hilbert action.
\end{abstract}

\maketitle

\tableofcontents

\setcounter{section}{0}
\section{Introduction}
In a previous paper \cite{cg:qgravity} we proved the quantization of gravity using canonical quantization to obtain a setting in which the standard techniques of $QFT$ can be applied to achieve a quantization of the gravitational field, i.e., the gravitational field can be treated like a non-gravitational field.

In order to make this approach work four new ideas had to be introduced in the process of canonical quantization:

\cvm
(i) We eliminated the diffeomorphism constraint by proving that it suffices to consider metrics that split according to
\begin{equation}
d\bar s^2=-w^2 (dx^0)^2+g_{ij}dx^idx^j
\end{equation}
after introducing a global time function $x^0$. The underlying spacetime $N=N^{n+1}$ can be considered to be a topological product
\begin{equation}
N=I\times\so
\end{equation}
where $I\su\R[]$ is an open interval, $\so$ a Cauchy hypersurface, fixed for all metrics under consideration, and $g_{ij}=g_{ij}(x^0,x)$, $x\in\so$, a Riemannian metric.

\cvm
(ii) The volume element $\sqrt g$, $g=\det(g_{ij})$, is a density and it appears explicitly in the Lagrangian and in the Hamiltonian. However, the Hamiltonian has to be an invariant, i.e., a function and not a density. To overcome this difficulty we fixed a metric $\chi\in T^{0,2}(\so)$ and defined the function $\f$ by
\begin{equation}
\f^2=\frac{\det(g_{ij})}{\det(\chi_{ij})}
\end{equation}
such that $\f=\f(x,g_{ij})$ and
\begin{equation}
\sqrt g=\f\sqrt\chi.
\end{equation}
The density $\sqrt\chi$ will be later ignored when performing the Legendre transformation in accordance with Mackey's advice to only use rectangular coordinates in canonical quantization, \cf \cite[p. 94]{mackey:book}.

\cvm
(iii) After the Legendre transformation the momenta depend on $x\in\so$. To overcome this difficulty we consider a fiber bundle with base space $\so$ where the fibers are the positive definite metrics $g_{ij}(x)$ over $x$, i.e., a fiber $F(x)$ is an open, convex cone in a finite dimensional vector space. We treat this cone as a manifold endowing it with the DeWitt metric which is Lorentzian. It turns out that $F(x)$ is globally hyperbolic. Let us call the bundle $E$. Each fiber has a Cauchy hypersurface $M(x)$ and we denote the corresponding bundle  by $\hat E$.

The introduction of the bundle $E$ simplifies the mathematical model after canonical quantization dramatically. The Hamiltonian operator $H$ is a normally hyperbolic differential operator acting only in the fibers which are globally hyperbolic spacetimes and the Wheeler-DeWitt equation is the hyperbolic equation
\begin{equation}
Hu=0,
\end{equation}
where $u$ is defined in $E$.

The Cauchy problem
\begin{equation}\lae{1.6}
\begin{aligned}
Hu&=f\\
\fv uM&=u_0\\
\fv{D_\nu u}M&=u_1
\end{aligned}
\end{equation}
is uniquely solvable in $E$ with $u\in C^\un(E,\K)$, $\K=\R[]\vee\K=\Cc$, for arbitrary
\begin{equation}
u_0,u_1\in C^\un_c(\hat E,\K)\q\wed\q f\in C^\un_c(E,\K).
\end{equation}

\cvm
(iv) In view of \re{1.6} the standard techniques of $QFT$, slightly modified to accept the present setting, can be applied to construct a quantum field $\F_{\hat E}$ which maps functions $u\in C^\un_c(E,\R[])$ to self-adjoint operators in the symmetric Fock space created from the Hilbert space
\begin{equation}
H_{\hat E}=L^2(\hat E,\Cc).
\end{equation}
The quantum field also satisfies the Wheeler-DeWitt equation in the distributional sense.

The quantization of the gravitational field makes it possible to obtain a unified quantum theory describing the interaction of gravity with other fundamental forces. In this paper we look at the interaction of gravity with a Yang-Mills and a Higgs field.

Let $N=N^{n+1}$ be globally hyperbolic spacetime with metric $(\bar g_{\al\bet})$, where the indices run from  $0\le \al,\bet\le n$, $\mc G$ a compact, semi-simple, connected Lie group with Lie algebra $\mf g$, and let  $E_1=(N,\mf g,\pi,\Ad(\mc G))$ be the corresponding adjoint bundle with base space $N$. Then we consider the functional
\begin{equation}
J=\al_N^{-1}\int_N(\bar R-2\Lam)+\int_N\{L_{YM}+L_H\},
\end{equation}
where $\al_N$ is a positive coupling constant, $\bar R$ the scalar curvature, $\Lam$ a cosmological constant, $L_{YM}$ the energy of a connection in $E_1$ and $L_H$ the energy of a Higgs field with values in $\mf g$. The integration over $N$ is to be understood symbolically, since we shall always integrate over an open precompact subset $\tilde\Om\su N$.

Let $A=(A^a_\mu)$ be a connection in $E_1$. We shall prove in \frt{3.3} that it will be sufficient to only consider connections satisfying the Hamilton gauge
\begin{equation}
A^a_0=0,
\end{equation}
thereby eliminating the Gau{\ss} constraint, such that the only remaining constraint is the Hamilton constraint.

Using the $ADM$ partition of $N$, \cf \cite{adm:old}, such that
\begin{equation}
N=I\times \so,
\end{equation}
where $\so$ is a Cauchy hypersurface and applying canonical quantization we obtain a Hamilton operator $H$ which is a normally hyperbolic operator in a fiber bundle $E$ with base space $\so$ and fibers
\begin{equation}
F(x)\times (\mf g\otimes T^{0,1}_x(\so))\times \mf g,\qq x\in \so,
\end{equation}
which are equipped with a Lorentzian metric. The Hamilton operator is only acting in the fibers and the Wheeler-DeWitt equation has the form
\begin{equation}
Hu=0,
\end{equation}
where $u\in C^\un(E,\Cc)$.

In \frt{5.1} we prove that the fibers are globally hyperbolic and have Cauchy hypersurfaces $M_0=M_0(x)$, $x\in\so$. Hence the results of \cite[Sections 5 \& 6]{cg:qgravity} are applicable leading to the existence of a quantum field $\F_{\hat E}$ which satisfies the Wheeler-DeWitt equation in the distributional sense and which maps $C^\un_c(E,\R[])$ to self-adjoint operators in the symmetric Fock space constructed from the Hilbert space
\begin{equation}
H_{\hat E}=L^2(\hat E,\Cc),
\end{equation}
where $\hat E$ is the bundle with base space $\so$ and fibers $M_0$. A more detailed analysis of applying the method of second quantization in the present situation can be found in \cite[Section 5]{cg:uqtheory2}.

\section{The Yang-Mills functional}

Let $N=N^{n+1}$ be a globally hyperbolic spacetime with metric $(\bar g_{\al\bet})$, $\mc G$ a compact, semi-simple, connected Lie group, $\mf g$ its Lie algebra and $E_1=(N,\mf g,\pi, \Ad(\mc G))$ the corresponding adjoint bundle with base space $N$. The Yang-Mills functional is then defined by
\begin{equation}\lae{3.1}
\begin{aligned}
J_{YM}&=\int_N-\tfrac14 F_{\mu\lam}F^{\mu\lam}=\int_N-\tfrac14 \ga_{ab}\bar g^{\mu\rho_2}\bar g^{\lam\rho_1}F^a_{\mu\rho_1}F^b_{\rho_2\lam},
\end{aligned}
\end{equation}
where $\ga_{ab}$ is the Cartan-Killing metric in $\mf g$,
\begin{equation}
F^a_{\mu\lam}=A^a_{\lam,\mu}-A^a_{\mu,\lam}+f^a_{bc}A^b_\mu A^c_\lam
\end{equation}
is the curvature of a connection
\begin{equation}
A=(A^a_\mu)
\end{equation}
in $E$ and
\begin{equation}\lae{3.4}
f_c=(f^a_{cb})
\end{equation}
are the structural constants of $\mf g$. The integration over $N$ is to be understood symbolically since we shall always integrate over an open precompact subset $\tilde\Om$ of $N$.

Let $A$ be a Yang-Mills connection, i.e., the first variation of $J_{YM}$ at the point $A$ vanishes with respect to compact variations of $A$, then $A$ satisfies the Euler-Lagrange equations
\begin{equation}
F^{a\mu}_{\lam\hp{\mu};\mu}=0,
\end{equation}
where we remind that covariant derivatives are always full tensors.

\bd\lad{3.1}
The adjoint bundle $E_1$ is vector bundle; let $E_1^*$ be the dual bundle, then we denote by 
\begin{equation}
T^{r,s}(E_1)=\C(\underset{r}{\underbrace{E_1\otimes\cdots\otimes E_1}}\otimes\underset{s}{\underbrace{E_1^*\otimes\cdots\otimes E_1^*}})
\end{equation}
the sections of the corresponding tensor bundle.
\ed
Thus, we have
\begin{equation}
F^a_{\mu\lam}\in T^{1,0}(E_1)\otimes T^{0,2}(N).
\end{equation}

When we fix a connection $\bar A$ in $E_1$, then a general connection $A$ can be written in the form
\begin{equation}\lae{3.8} 
A^a_\mu=\bar A^a_\mu+\tilde A^a_\mu,
\end{equation}
where $\tilde A^a_\mu$ is a tensor
\begin{equation}
\tilde A^a_\mu\in T^{1,0}(E_1)\otimes T^{0,1}(N).
\end{equation}
To be absolutely precise a connection in $E_1$ is of the form
\begin{equation}
f_cA^c_\mu,
\end{equation}
where $f_c$ is defined in \re{3.4}; $A^a_\mu$ is merely a coordinate representation.
\bd
A connection $A$ of the form \re{3.8} is sometimes also denoted by $(\bar A^a_\mu,\tilde A^a_\mu)$.
\ed
Since we assume that there exists a globally defined time function $x^0$ in $N$ we may consider globally defined tensors $(\tilde A^a_\mu)$ satisfying
\begin{equation}\lae{3.11} 
\tilde A^a_0=0. 
\end{equation}
These tensors can be written in the form $(\tilde A^a_i)$ and they can be viewed as maps
\begin{equation}
(\tilde A^a_i):N\ra \mf g\otimes T^{0,1}(\so),
\end{equation}
where $\so$ is a Cauchy hypersurface of $N$, e.g.,  a coordinate slice
\begin{equation}
\so=\{x^0=\const\}.
\end{equation}
 It is well-known that the Yang-Mills Lagrangian is singular and requires a local gauge fixing when applying canonical quantization. We impose a local gauge fixing by stipulating that the connection $\bar A$ satisfies
 \begin{equation}\lae{3.14}
\bar A^a_0=0.
\end{equation}
Hence, all connections in \re{3.8} will obey this condition since we also stipulate that the tensor fields $\tilde A^a_\mu$ have vanishing temporal components as in \re{3.11}. The gauge \re{3.14} is known as the \tit{Hamilton gauge}, \cf \cite[p. 82]{faddeev:book}. However, this gauge fixing leads to the so-called Gau{\ss} constraint, since the first variation in the class of these connections will not yield the full Yang-Mills equations.

In the following theorem we shall prove that the Gau{\ss} constraint does not exist and  that it suffices to consider connections of the form \re{3.8} satisfying \re{3.11} and \re{3.14} in the Yang-Mills functional $J_{YM}$.
\bt\lat{3.3}
Let  $\tilde \Om\Su N$ be open and precompact such that there exists a local trivialization of $E_1$ in $\tilde \Om$. Let $A=(\bar A^a_\mu,\tilde A^a_\mu)$ be a connection satisfying \re{3.11} and \re{3.14} in $\tilde \Om$, and suppose that the first variation of $J_{YM}$ vanishes at $A$ with respect to compact variations of $\tilde A^a_\mu$ all satisfying \re{3.11}. Then $A$ is a Yang-Mills connection, i.e., the  Yang-Mills equation
\begin{equation}
F^{a\mu}_{\lam\hp{\mu};\mu}=0
\end{equation}
is valid in $\tilde \Om$. 
\et
\bp
Let $\h^a_\mu$ be an arbitrary tensor field with compact support in $\tilde\Om$ satisfying
\begin{equation}
\h^a_0=0
\end{equation}
and define the connections
\begin{equation}\lae{3.17}
A(\e)=(\bar A^a_\mu,\tilde A^a_\mu+\e\h^a_\mu).
\end{equation}
Differentiating the functional
\begin{equation}\lae{3.18}
J_{YM}(\e)=\int_{\tilde\Om}-\tfrac14 F_{\mu\lam}(\e)F^{\mu\lam}(\e)
\end{equation}
with respect to $\e$ and evaluating in $\e=0$ yields
\begin{equation}\lae{3.19}
\df{J_{YM}}\e=-\int_{\tilde\Om}\ga_{ab}F^{a\mu\lam}\h^b_{\lam;\mu}=\int_{\tilde\Om}\ga_{ab}F^{a\mu\lam}_{\hp{a\mu\lam};\mu}\h^b_\lam.
\end{equation}
Assuming that the first variation  of the functional vanishes we deduce
\begin{equation}\lae{3.20}
F^{ai\mu}_{\hp{ai\m};\mu}=0
\end{equation}
which is equivalent to
\begin{equation}
F^{a\hp{i}\mu}_{\hp{a}i\hp{\mu};\mu}=0
\end{equation}
since we only consider spacetime metrics $(\bar g_{\al\bet})$ that splits, i.e.,
\begin{equation}\lae{3.22}
d\bar s^2=-w^2(dx^0)^2+g_{ij}(x^0,x) dx^idx^j
\end{equation}
in view of the results in \cite[Theorem 3.2]{cg:qgravity}. Similarly, the conditions
\begin{equation}
F^{a0\mu}_{\hp{a0\m};\mu}=0
\end{equation}
and
\begin{equation}
F^{a\hp{0}\mu}_{\hp{a}0\hp{\mu};\mu}=0
\end{equation}
are equivalent.

To prove that $A$ also satisfies
\begin{equation}
F^{0\mu}_{\hp{0\mu};\mu}=0
\end{equation}
in $\tilde \Om$, we argue by contradiction supposing there exists $(t_0,x_0)\in\tilde\Om$ such that
\begin{equation}
F^{0\mu}_{\hp{0\mu};\mu}(t_0,x_0)\not=0.
\end{equation}

Define
\begin{equation}
\xi^a=F^{a0\mu}_{\hp{a0\mu};\mu}\msp[1]\bar g_{00}
\end{equation}
so that
\begin{equation}
\ga_{ab}\xi^aF^{b0\mu}_{\hp{b0\mu};\mu}<0
\end{equation}
in $(t_0,x_0)$. Choosing a cut-off function $\f=\f(t,x)$ satisfying $\f(t_0,x_0)=1$ we then infer
\begin{equation}\lae{3.29}
\ga_{ab}\tilde\xi^aF^{b0\mu}_{\hp{b0\mu};\mu}\le0
\end{equation}
in $N$ and strictly negative in $(t_0,x_0)$, where
\begin{equation}
\tilde\xi^a=\xi^a\f.
\end{equation}

Next we consider the gauge transformation $\om=\om(t,x)$ where $\om$ is the flow
\begin{equation}
\begin{aligned}
\dot \om&=-\om\e f_c\tilde\xi^c,\\
\om(t_0,x)&=\id,
\end{aligned}
\end{equation}
which is well defined in a neighbourhood of $\supp\f$. After the gauge transformation the connections $A(\e)$ in \re{3.17} look like
\begin{equation}
\om f_c A^c_\mu(\e)\om^{-1}-\om_\mu\om^{-1}
\end{equation}
and the component $\mu=0$ is equal to
\begin{equation}
-\dot \om \om^{-1}=\e \om f_c\tilde\xi^c\om^{-1}.
\end{equation}
Since the Yang-Mills functional is gauge invariant its first variation still vanishes after the gauge transformation and we deduce from \re{3.19} and \re{3.20}
\begin{equation}
0=\int_{\tilde\Om}\ga_{ab}F^{a0\mu}_{\hp{a0\mu};\mu}\tilde\xi^b
\end{equation}
contradicting \re{3.29}.
\ep
\br
Gauge fixing is an appropriate method for reducing the number of independent variables, but in the context of canonical quantization it is only legitimate if it is also used before deriving the Euler-Lagrange equation and if in addition it is proved that the correct Euler-Lagrange equation is still valid.
\er
Let $(B_{\rho_k}(x_k))_{k\in\N}$ be a covering of $\so$ by small open balls such that each ball lies in a coordinate chart of $\so$. Then the cylinders
\begin{equation}
U_k=I\times B_{\rho_k}(x_k)
\end{equation}
are a covering of $N$ such that each $U_k$ is contractible, hence each bundle $\pi^{-1}(U_k)$ is trivial and the connection $\bar A$ can be expressed in coordinates in each $U_k$
\begin{equation}
\bar A=(\bar A^a_\mu)=f_aA^a_\mu dx^\mu.
\end{equation}
We shall prove:
\bl\lal{2.5}
In each cylinder $U_k$ there exists a gauge transformation $\om=\om(t,x)$ such that
\begin{equation}
\bar A^a_0(t,x)=0\qq\A\, (t,x)\in U_k
\end{equation}
after applying the gauge transformation.
\el
\bp
For fixed $k$ we consider the flow
\begin{equation}
\begin{aligned}
\dot\om&=\om f_c\bar A^c_0,\\
\om(0,x)&=\id,\qq \qq  x\in B_{\rho_k}(x_k).
\end{aligned}
\end{equation}
For fixed $x\in B_{\rho_k}(x_k)$ the integral curve exists on a maximal interval $J_x$. If we can show $J_x=I$, then the lemma is proved.

The claim is obvious, since the integral curve cannot develop singularities, for let $\spd\cdot\cdot$ be the negative of the Killing metric, then
\begin{equation}
\begin{aligned}
\spd{\dot\om}{\dot\om}&=-\tr(\om A_0 \om A_0)\\
&=-\tr(A_0A_0)=\ga_{ab}A^a_0A^b_0
\end{aligned}
\end{equation}
from which the result immediately follows.
\ep
\bl\lal{2.6}
Let $U_k$, $U_l$ be overlapping cylinders and let $\om=\om(t,x)$ be a gauge transformation relating the respective representations of the connection $\bar A$ in the overlap $U_k\ii U_l$ where both representations use the Hamilton gauge, then $\om$ does not depend on $t$, i.e.,
\begin{equation}
\dot\om=0.
\end{equation}
\el
\bp
Let $(\hat{\bar A}^a_\mu)$ \resp $(\bar A^a_\mu)$ be the representations of $\bar A$ in $U_k$ \resp $U_l$ such that
\begin{equation}
\hat{\bar A}^a_0=\bar A^a_0=0,
\end{equation}
then
\begin{equation}
\hat{\bar A}_0=\om \bar A_0\om^{-1}-\dot\om \om^{-1} ,
\end{equation}
hence
\begin{equation}
\dot\om=0\qq \text{in}\q U_k\ii U_l.
\end{equation}
\ep

Let $E_0$ be the adjoint bundle
\begin{equation}
E_0=(S_0,\mf g,\pi,\Ad(\mc G))
\end{equation}
with base space $\so$, where the gauge transformations only depend on the spatial variables $x=(x^i)$. For fixed $t$ $A^a_{i,0}$ are elements of $T^{1,0}(E_0)\otimes T^{0,1}(\so)$
\begin{equation}
A^a_{i,0}\in T^{1,0}(E_0)\otimes T^{0,1}(\so),
\end{equation}
but the vector potentials $A^a_i(t,\cdot)$ are connections in $E_0$ for fixed $t$ and therefore cannot be used as independent variables, since the variables should be the components of a tensor. However, in view of the results in \rl{2.5} and \rl{2.6} the difference
\begin{equation}
\tilde A^a_i(t,\cdot)=A^a_i(t,\cdot)-\bar A^a_i(0,\cdot)\in T^{1,0}(E_0)\otimes T^{0,1}(\so).
\end{equation}
Hence, we shall define $\tilde A^a_i$ to be the independent variables such that
\begin{equation}
A^a_i=\bar A^a_i(0,\cdot)+\tilde A^a_i
\end{equation}
and we infer
\begin{equation}
A^a_{i,0}=\tilde A^a_{i,0}.
\end{equation}
In the Hamilton gauge we therefore have
\begin{equation}
F^a_{0i}=\tilde A^a_{i,0}
\end{equation}
and hence we conclude
\begin{equation}
-\tfrac14 F_{\mu\lam}F^{\mu\lam}=\tfrac12 w^{-2}g^{ij}\ga_{ab}\tilde A^a_{i,0}\tilde A^b_{j,0}-\tfrac14 F_{ij}F^{ij},
\end{equation}
where we used \re{3.22}.

Writing the density
\begin{equation}
\sqrt g=\sqrt{\det g_{ij}}
\end{equation}
in the form
\begin{equation}\lae{3.38} 
\sqrt g=\f \sqrt{\det \chi_{ij}},
\end{equation}
where $\chi$ is a fixed Riemannian metric in $S_0$, $\chi_{ij}=\chi_{ij}(x)$, such that $0<\f=\f(x,g_{ij})$ is a function, we obtain as Lagrangian function
\begin{equation}\lae{3.39}
L_{YM}=\tfrac12\ga_{ab}g^{ij}\tilde A^a_{i,0}\tilde A^b_{j,0}w^{-1}\f-\tfrac14 F_{ij}F^{ij}w\f.
\end{equation}
The $\tilde A^a_i(t,\cdot)$ can be looked at to be mappings from $N$ to $T^{1,0}(E_0)\otimes T^{0,1}(\so)$
\begin{equation}
\tilde A^a_i: N\ra T^{1,0}(E_0)\otimes T^{0,1}(\so).
\end{equation}

The fibers of $T^{1,0}(E_0)\otimes T^{0.1}(\so)$ are the tensor products
\begin{equation}
\mf g\otimes T^{0,1}_x(\so),\qq x\in \so,
\end{equation}
which are vector spaces equipped with metric
\begin{equation}
\ga_{ab}\otimes g^{ij}. 
\end{equation}
For our purposes it is more convenient to consider the fibers to be Riemannian manifolds endowed with the above metric. Let $(\zeta^p)$, $1\le p\le n_1n$, where $n_1=\dim\mf g$, be  local coordinates and
\begin{equation}
(\zeta^p)\ra \tilde A^a_i(\zeta^p)\equiv \tilde A(\zeta)
\end{equation}
be a local embedding, then the metric has the coefficients
\begin{equation}
G_{pq}=\spd{\tilde A_p}{\tilde A_q}=\ga_{ab}g^{ij}\tilde A^a_{i,p}\tilde A^b_{j,q}.
\end{equation}
Hence, the Lagrangian $L_{YM}$ in \re{3.39} can be expressed in the form
\begin{equation}
L_{YM}=\tfrac12G_{pq}\dot\zeta^p\dot\zeta^qw^{-1}\f-\tfrac14F_{ij}F^{ij}w\f
\end{equation}
and we deduce
\begin{equation}
\tilde\pi_p=\pde{L_{YM}}{\dot\zeta^p}=G_{pq}\dot\zeta^qw^{-1}\f
\end{equation}
yielding the Hamilton function
\begin{equation}
\begin{aligned}
\hat H_{YM}&=\pi_p\dot\zeta^p-L_{YM}\\
&=\tfrac12 G_{pq}(\dot\zeta^pw^{-1}\f)(\dot\zeta^qw^{-1}\f)w\f^{-1}+\tfrac14F_{ij}F^{ij}w\f\\
&=\tfrac 12G^{pq}\tilde\pi_p\tilde\pi_qw\f^{-1}+\tfrac14F_{ij}F^{ij}w\f\\
&\equiv H_{YM}w.
\end{aligned}
\end{equation}
Thus, the Hamiltonian that will enter the Hamilton constraint equation is
\begin{equation}\lae{3.51}
H_{YM}=\tfrac 12\f^{-1}G^{pq}\tilde\pi_p\tilde\pi_q+\tfrac14F_{ij}F^{ij}\f.
\end{equation}

\section{The Higgs functional}
Let $\F$ be a scalar field, a map from $N$ to $E_1$,
\begin{equation}
\F:N\ra E_1,
\end{equation}
i.e., $\F$ is a section of $E_1$. Using the notation in \frd{3.1}, we also write
\begin{equation}
\F\in T^{1,0}(E_1).
\end{equation}
The Higgs Lagrangian is defined by
\begin{equation}\lae{4.3}
L_H=-\tfrac12\bar g^{\al\bet}\ga_{ab}\F^a_\al\F^b_\bet-V(\F),
\end{equation}
where $V$ is a smooth potential. We assume that in a local coordinate system $\F$ has real coefficients. The covariant derivatives of $\F$ are defined by a connection $A=(A^a_\mu)$ in $E_1$ 
\begin{equation}
\F^a_\mu=\F^a_{,\mu}+f^a_{cb}A^c_\mu\F^b.
\end{equation}
As in the preceding section we work in a local trivialization of $E_1$ using the Hamilton gauge, i.e.,
\begin{equation}
A^a_0=0,
\end{equation}
hence, we conclude
\begin{equation}
\F^a_0=\F^a_{,0}.
\end{equation}

Expressing the density $g$ as in \fre{3.38} we obtain Lagrangian
\begin{equation}
L_H=\tfrac12 \ga_{ab}\F^a_{,0}\F^b_{,0}w^{-1}\f-\tfrac12g^{ij}\ga_{ab}\F^a_i\F^b_jw\f-V(\F)w\f
\end{equation}
which we have to use for the Legendre transformation. Before applying the Legendre transformation we again consider the vector space $\mf g$ to be a Riemannian manifold with metric $\ga_{ab}$. The representation of $\F$ in the form $(\F^a)$ can be looked at to be the representation in a local coordinate system $(\Theta^a)$.

Let us define
\begin{equation}
p_a=\pde{L_H}{\dot\F^a},\qq\dot\F^a=\F^a_{,0},
\end{equation}
then we obtain the Hamiltonian
\begin{equation}
\begin{aligned}
\hat H_H&=p_a\dot\F^a-L_H\\
&=\tfrac12\ga_{ab}(\dot\F^aw^{-1}\f)(\dot\F^bw^{-1}\f)w\f^{-1}+\tfrac12 g^{ij}\ga_{ab}\F^a_i\F^b_jw\f+V(\F)w\f\\
&\equiv H_Hw.
\end{aligned}
\end{equation}
Thus, the Hamiltonian which will enter the Hamilton constraint is
\begin{equation}\lae{4.10}
H_H=\tfrac12\f^{-1}\ga^{ab}p_ap_b+\tfrac12 g^{ij}\ga_{ab}\F^a_i\F^b_j\f+V(\F)\f
\end{equation}

\section{The Wheeler-DeWitt equation}
The interaction of gravity with the Yang-Mills  and the Higgs field is described by the functional
\begin{equation}\lae{5.1}
J=\al_N^{-1}\int_{\tilde\Om}(\bar R-2\Lam)+\int_{\tilde\Om}\{L_{YM}+L_H\},
\end{equation}
where $\tilde\Om\Su N$ is an open precompact set, $\bar R$ the scalar curvature, $\Lam$ a cosmological constant and $L_{YM}$ \resp $L_H$ the Lagrangians in \fre{3.1} \resp \fre{4.3}.

As we proved in \cite{cg:qgravity} we may only consider metrics $\bar g_{\al\bet}$ that split with respect to some fixed globally defined time function $x^0$ such that
\begin{equation}
d\bar s^2=-w^2 (dx^0)^2+g_{ij}dx^idx^j
\end{equation}
where $g(x^0,\cdot)$ are Riemannian metrics in $\so$,
\begin{equation}
\so=\{x^0=0\}.
\end{equation}
The first functional on the right-hand side of \re{5.1} can be written in the form 
\begin{equation}\lae{5.4}
\al^{-1}_N\int_a^b\int_\Om\{\tfrac14G^{ij,kl}\dot g_{ij}\dot g_{kl}+R-2\Lam\}w\f,
\end{equation}
where
\begin{equation}
G^{ij,kl}=\tfrac12\{g^{ik}g^{jl}+g^{il}g^{jk}\}-g^{ij}g^{kl}
\end{equation}
is the DeWitt metric,
\begin{equation}
(g^{ij}=(g_{ij})^{-1},
\end{equation}
$R$ the scalar curvature of the slices
\begin{equation}
\{x^0=t\}
\end{equation}
with respect to the metric $g_{ij}(t,\cdot)$, and where we also assumed that $\tilde\Om$ is a cylinder
\begin{equation}
\tilde\Om=(a,b)\times\Om,\qq\Om\Su \so,
\end{equation}
and, where now, we  also assume that $\Om\su U_k$ for some $k\in \N$.

The Riemannian metrics $g_{ij}(t,\cdot)$ are elements of the bundle $T^{0,2}(\so)$. Denote by $\tilde E$ the fiber bundle with base $\so$ where the fibers $F(x)$ consists of the Riemannian metrics $(g_{ij})$. We shall consider each fiber to be a Lorentzian manifold equipped with the DeWitt metric. Each fiber $F$ has dimension
\begin{equation}
\dim F=\frac{n(n+1)}2\equiv m+1.
\end{equation}
Let $(\xi^r)$, $0\le r\le m$, be  coordinates for a local trivialization such that
\begin{equation}
g_{ij}(x,\xi^r)
\end{equation}
is a local embedding. The DeWitt metric is then expressed as
\begin{equation}
G_{rs}=G^{ij,kl}g_{ij,r}g_{kl,s},
\end{equation}
where a comma indicates partial differentiation.  
In the new coordinate system the curves 
\begin{equation}
t\ra g_{ij}(t,x)
\end{equation}
can be written in the form
\begin{equation}
t\ra \xi^r(t,x)
\end{equation}
and we infer
\begin{equation}
G^{ij,kl}\dot g_{ij}\dot g_{kl}=G_{rs}\dot\xi^r\dot\xi^s. 
\end{equation}
Hence, we can express \re{5.4} as 
\begin{equation}\lae{3.49}
J=\int_a^b\int_\Om \al_n^{-1}\{\tfrac14 G_{rs}\dot\xi^r\dot\xi^sw^{-1}\f+(R-2\Lam)w\f\},
\end{equation}
where we now refrain from writing down the density $\sqrt\chi$ explicitly, since it does not depend on $(g_{ij})$ and therefore should not be part of the Legendre transformation.  Here we follow Mackey's advice in \cite[p. 94]{mackey:book} to always consider rectangular coordinates when applying canonical quantization, which can be rephrased that the Hamiltonian has to be a coordinate invariant, hence no densities are allowed.

Denoting the Lagrangian \tit{function} in \re{3.49} by $L$, we define
\begin{equation}
\pi_r= \pde L{\dot\xi^r}=\f G_{rs}\frac1{2\al_N}\dot\xi^sw^{-1}
\end{equation}
and we obtain for the Hamiltonian function $\hat H_G$
\begin{equation}
\begin{aligned}
\hat H_G&=\dot\xi^r\pde L{\dot\xi^r}-L\\
&=\f G_{rs}\big(\frac1{2\al_N}\dot\xi^rw^{-1}\big)\big(\frac1{2\al_N}\dot\xi^sw^{-1}\big) w\al_N-\al_N^{-1}(R-2\Lam)\f w\\
&=\f^{-1}G^{rs}\pi_r\pi_s w\al_N-\al^{-1}_N(R-2\Lam)\f w\\
&\equiv H_G w,
\end{aligned}
\end{equation}
where $G^{rs}$ is the inverse metric. Hence,
\begin{equation}\lae{5.18}
H_G=\al_N\f^{-1}G^{rs}\pi_r\pi_s-\al_N^{-1}(R-2\Lam)\f
\end{equation}
is the Hamiltonian that will enter the Hamilton constraint.

Combing the three Hamilton functions in \fre{3.51}, \fre{4.10} and \re{5.18} the Hamilton constraint has the form
\begin{equation}
H=H_G+H_{YM}+H_H=0,
\end{equation}
where
\begin{equation}
H=H(\xi^r,\zeta^p,\Theta^a,\pi_s,\tilde\pi_q,p_b).
\end{equation}
Here $(\xi^r,\zeta^p,\Theta^a)$ are local sections of a bundle $E$ with base space $\so$ and fibers
\begin{equation}\lae{5.21}
F(x)\times (\mf g\otimes T^{01}_x(\so))\times \mf g
\end{equation}
where the fibers are Riemannian manifolds endowed with product metric
\begin{equation}\lae{5.22}
G=\f\diag(\al_N^{-1}G_{rs},2G_{pq},2\ga_{ab}).
\end{equation}
Applying quantization, by setting $\hbar=1$, we replace
\begin{equation}
\pi_r=\pi_r(x)\ra \frac1i\pde{}{\xi^r(x)}
\end{equation}
and similarly for the other conjugate momenta $\tilde \pi_p$ and $p_a$.

After quantization we obtain a normally hyperbolic differential operator, which we shall also denote by $H$, acting only in the fibers and the Wheeler-DeWitt equation looks like
\begin{equation}
Hu=0,
\end{equation}
where $u\in C^\un(E,\Cc)$.

The fibers are Lorentzian manifolds equipped with the Lorentz metric $G$. If we can prove that the fibers are globally hyperbolic, then the techniques of $QFT$, appropriately modified to work in the bundle, can be applied to construct a quantum field $\F_{\hat E}$ which maps functions $u\in C^\un(E,\R[])$ to essentially self-adjoint operators in the symmetric Fock space created from the Hilbert space 
\begin{equation}
H_{\hat E}=L^2(\hat E,\Cc).
\end{equation}
$\hat E$ is the bundle with base space $\so$ and fibers $M_0$, where $M_0(x)$ is a Cauchy hypersurface in the corresponding fiber \re{5.21} in $E$, \cf \cite[Section 6]{cg:qgravity} for details.

The Lorentzian nature of $G$ is due to the metric $G_{rs}$ which is the DeWitt metric. In \cite[Section 4]{cg:qgravity} we proved that
\begin{equation}
\tau=\log\f
\end{equation}
is a time function and that the hypersurface
\begin{equation}
M=\{\f=1\}=\{\tau=0\}
\end{equation}
is a Cauchy hypersurface in $F(x)$, hence $F(x)$ is globally hyperbolic, \cf \cite[Corollary 39, p. 422]{bn}.

We shall prove:
\bt\lat{5.1}
The hypersurfaces $M_0=M_0(x)$, $x\in \so$, 
\begin{equation}
M_0=M\times (\mf g\otimes T^{0,1}_x(\so))\times\mf g
\end{equation}
are Cauchy hypersurfaces in each fiber over $x\in\so$.
\et
\bp
We follow the proof in \cite[Lemma 4.3]{cg:qgravity}. Fix $x\in \so$, then the metric $G_{rs}$ splits and can be expressed in the form
\begin{equation}
ds^2=c\{-d\tau^2+G_{AB}d\xi^Ad\xi^B\},
\end{equation}
where $c$ is a positive constant,
\begin{equation}\lae{5.29}
\tau=\xi^0\q\wed\q -\un<\tau<\un,
\end{equation}
and $(\xi^A)$, $1\le A\le m$, are local coordinates for $M$. The metric $G_{AB}$ is the metric of the hypersurface $M$ when the ambient space $F(x)$ is equipped with the DeWitt metric; $G_{AB}$ does not depend on $\tau$.

Let $\ga=\ga(s)$, $s\in I$, be an inextendible future directed timelike curve in $F$. We have to prove that it intersects with $M$ exactly once. It suffices to show that it intersects $M$, the uniqueness is trivial.

Suppose that $\ga$ does not intersect $M$.  Assume there exists $s_0\in I$ such that
\begin{equation}
\tau(\ga(s_0))<0
\end{equation}
and assume from now on that $s_0$ is the left endpoint of $I$. Since $\tau$ is continuous the whole curve $\ga$ must be contained in the past of $M$. 

From the relation \re{5.29} we deduce that the whole metric $\f^{-1}G$ in \re{5.22} splits according to \re{5.29}. Stipulating that $(\xi^0,\xi^A)$ represent a coordinate system for the fiber in \re{5.21} and not just for the $F(x)$ component, we may consider \re{5.29} to represent the metric of the whole fiber.

Hence, $\ga=(\ga^0,\ga^A)$ and because $\ga$ is timelike we deduce
\begin{equation}
G_{AB}\dot\ga^A\dot\ga^B\le \abs{\dot\ga^0}^2
\end{equation}
and thus
\begin{equation}
\sqrt{G_{AB}\dot\ga^A\dot\ga^B}\le \dot\ga^0,
\end{equation}
since $\ga$ is future directed. Let
\begin{equation}
\tilde\ga=(\ga^A)
\end{equation}
be the projection of $\ga$ to $M_0$, then the length of $\tilde \ga$ is bounded
\begin{equation}
L(\tilde\ga)=\int_I\sqrt{G_{AB}\dot\ga^A\dot\ga^B}\le \int_I\dot\ga^0\le-\ga^0(s_0).
\end{equation}

Let us express the curve $\tilde\ga$ in the original coordinate system
\begin{equation}
\tilde\ga=(g_{ij}(x,s),\tilde A^a_k(x,s),\Theta^b(x,s)),
\end{equation}
then
\begin{equation}
\begin{aligned}
G_{AB}\dot\ga^A\dot\ga^B&=g^{ik}g^{jl}\dot g_{ij}\dot g_{kl}+\ga_{ab}g^{kl}\dot {\tilde A}^a_k\dot {\tilde A}^b_l+\ga_{cd}\dot\Theta^c\dot\Theta^s,
\end{aligned}
\end{equation}
where we used that
\begin{equation}
g^{ij}\dot g_{ij}=0,
\end{equation}
since the normal to $M$ is a multiple of $g^{ij}$, and we conclude
\begin{equation}
\int_I\sqrt{g^{ik}g^{jl}\dot g_{ij}\dot g_{kl}}\le -\ga^0(s_0)
\end{equation}
and identical estimates for the other components. In \cite[Lemma 4.3]{cg:qgravity} we have shown that the metrics $g_{ij}(s)$ are uniformly equivalent and stay in a compact subset of $M$. Hence, we can replace the norm
\begin{equation}
\ga_{ab}g^{kl}\dot {\tilde A}^a_k\dot {\tilde A}^b_l
\end{equation}
by
\begin{equation}
\ga_{ab}\chi^{kl}\dot {\tilde A}^a_k\dot {\tilde A}^b_l,
\end{equation}
where $\chi_{ij}=\chi_{ij}(x)$ is a metric independent of $s$, and we conclude that the components ${\tilde A}^a_k$ are uniformly bounded and therefore contained a compact subset, since the ambient space is a finite dimensional vector space. 

The same argument is also valid in case of the components $\Theta^a(s)$ and we finally obtain a contradiction because an inextendible timelike curve cannot stay in a compact subset.
\ep

\bibliographystyle{hamsplain}
\providecommand{\bysame}{\leavevmode\hbox to3em{\hrulefill}\thinspace}
\providecommand{\href}[2]{#2}



\end{document}